\documentclass[useAMS,usegraphicx,usenatbib]{mn2e}          

\usepackage{times}

\newcommand{\Msun}{\ensuremath{\,{\rm M}_\odot}}                  
\newcommand{\Rsun}{\ensuremath{\,{\rm R}_\odot}}                  
\newcommand{\Teff}{\ensuremath{T_{\rm eff}}}                      
\newcommand{\Mjup}{\ensuremath{\,{\rm M}_{\rm Jup}}}              
\newcommand{\Rjup}{\ensuremath{\,{\rm R}_{\rm Jup}}}              
\newcommand{\Teq}{\ensuremath{T_{\rm eq}^{\,\prime}}}             
\newcommand{\safronov}{\ensuremath{\Theta}}                       
\newcommand{\kms}{\,km\,s$^{-1}$}                                 
\newcommand{\ms}{\,m\,s$^{-1}$}                                   
\newcommand{\mss}{\,m\,s$^{-2}$}                                  
\newcommand{\as}{\ensuremath{^{\prime\prime}}}                    
\newcommand{\FeH}{\ensuremath{\left[\frac{\rm Fe}{\rm H}\right]}} 
\newcommand{\pjup}{\ensuremath{\,\rho_{\rm Jup}}}                 
\newcommand{\psun}{\ensuremath{\,\rho_\odot}}                     
\newcommand{\mc}[1]{\multicolumn{2}{c}{#1}}
\newcommand{\mcc}[1]{\multicolumn{3}{c}{#1}}

\newcommand{\erc}[3]{\mc{\ensuremath{#1^{+#2}_{-#3}}}}
\newcommand{\erm}[3]{\mc{\ensuremath{#1^{+#2}_{-#3}}}}

\newcommand{\ermcc}[5]{\mcc{\ensuremath{{#1\,^{+#2}_{-#3}}\,^{+#4}_{-#5}}}}

\title[High-precision defocussed photometry of WASP-2]
      {High-precision photometry by telescope defocussing. III. The transiting planetary system WASP-2\thanks{Based on data collected by MiNDSTEp with the Danish 1.54\,m telescope at the ESO La Silla Observatory}}

\author[Southworth et al.]
       {John Southworth\,$^{1}$,                                                
        L.\ Mancini\,$^{2,3,4,5}$,                                              
        S.\ Calchi Novati\,$^{3,4,5}$,                                          
        M.\ Dominik\,$^{6}$\thanks{Royal Society University Research Fellow},   
        M.\ Glitrup\,$^{7}$,                                                    
        \newauthor
        T.\ C.\ Hinse\,$^{8,9}$,                                                
        U.\ G.\ J{\o}rgensen\,$^{9}$,                                           
        M.\ Mathiasen\,$^{9}$,                                                  
        D.\ Ricci\,$^{10}$,                                                     
        G.\ Maier\,$^{11}$,                                                     
        F.\ Zimmer\,$^{11}$,                                                    
        \newauthor
        V.\ Bozza\,$^{3,4,5}$,                                                  
        P.\ Browne\,$^{6}$,                                                     
        I.\ Bruni\,$^{12}$,                                                     
        M.\ Burgdorf\,$^{13}$,                                                  
        M.\ Dall'Ora\,$^{14}$,                                                  
        F.\ Finet\,$^{10}$,                                                     
        \newauthor
        K.\ Harps{\o}e\,$^{9}$,                                                 
        M.\ Hundertmark\,$^{15}$,                                               
        C.\ Liebig\,$^{11}$,                                                    
        S.\ Rahvar\,$^{16}$,                                                    
        G.\ Scarpetta\,$^{3,4,5}$,                                              
        J.\ Skottfelt\,$^{9}$,                                                  
        \newauthor
        B.\ Smalley\,$^{1}$,                                                    
        C.\ Snodgrass\,$^{17,18}$,                                              
        J.\ Surdej\,$^{10}$                                                     
        \\
        $^{1}$\,Astrophysics Group, Keele University, Newcastle-under-Lyme, ST5 5BG, UK \\
        $^{2}$\,Dipartimento di Ingegneria, Universit\`a� del Sannio, Corso Garibaldi 107, 82100-Benevento, Italy \\
        $^{3}$\,Dipartimento di Fisica ``E. R. Caianiello'', Universit\`a di Salerno, Via Ponte Don Melillo, 84084-Fisciano (SA), Italy \\
        $^{4}$\,Istituto Nazionale di Fisica Nucleare, Sezione di Napoli, Italy \\
        $^{5}$\,Istituto Internazionale per gli Alti Studi Scientifici (IIASS), Vietri Sul Mare (SA), Italy \\
        $^{6}$\,SUPA, University of St Andrews, School of Physics \& Astronomy, North Haugh, St Andrews, KY16 9SS, UK \\
        $^{7}$\,Department of Physics \& Astronomy, Aarhus University, Ny Munkegade, 8000 Aarhus C, Denmark \\
                $^{8}$\,Armagh Observatory, College Hill, Armagh, BT61 9DG, Northern Ireland, UK \\
        $^{9}$\,Niels Bohr Institute and Centre for Star and Planet Formation, University of Copenhagen, Juliane Maries vej 30, 2100 Copenhagen \O, Denmark \\
        $^{10}$\,Institut d'Astrophysique et de G\'eophysique, Universit\'e de Li\`ege, 4000 Li\`ege, Belgium \\
        $^{11}$\,Astronomisches Rechen-Institut, Zentrum f\"ur Astronomie, Universit\"at Heidelberg, M\"onchhofstra{\ss}e 12-14, 69120 Heidelberg, Germany \\
        $^{12}$\,INAF -- Osservatorio Astronomico di Bologna, Via Ranzani 1, 40127 Bologna, Italy \\
        $^{13}$\,Deutsches SOFIA Institut, NASA Ames Research Center, Mail Stop 211-3, Moffett Field, CA 94035, USA \\
        $^{14}$\,INAF -- Osservatorio Astronomico di Capodimonte, Via Moiarello 16, 80131 Napoli, Italy \\
        $^{15}$\,Institut f\"ur Astrophysik, Georg-August-Universit\"at G\"ottingen, Friedrich-Hund-Platz 1, 37077 G\"ottingen, Germany \\
        $^{16}$\,Department of Physics, Sharif University of Technology, Tehran, Iran \\
        $^{17}$\,European Southern Observatory, Casilla 19001, Santiago 19, Chile \\
        $^{18}$\,Max-Planck-Institute for Solar System Research, Max-Planck Str.\ 2, 37191 Katlenburg-Lindau, Germany
        }

\begin{document} \maketitle 

\begin{abstract}
We present high-precision photometry of three transits of the extrasolar planetary system WASP-2, obtained by defocussing the telescope, and achieving point-to-point scatters of between 0.42 and 0.73 mmag. These data are modelled using the {\sc jktebop} code, and taking into account the light from the recently-discovered faint star close to the system. The physical properties of the WASP-2 system are derived using tabulated predictions from five different sets of stellar evolutionary models, allowing both statistical and systematic errorbars to be specified. We find the mass and radius of the planet to be $M_{\rm b} = 0.847 \pm 0.038 \pm 0.024$\Mjup\ and $R_{\rm b} = 1.044 \pm 0.029 \pm 0.015$\Rjup. It has a low equilibrium temperature of $1280 \pm 21$\,K, in agreement with a recent finding that it does not have an atmospheric temperature inversion. The first of our transit datasets has a scatter of only 0.42\,mmag with respect to the best-fitting light curve model, which to our knowledge is a record for ground-based observations of a transiting extrasolar planet.
\end{abstract}



\section{Introduction}

\begin{table*} \centering
\caption{\label{tab:obslog} Log of the observations presented in this work. $N_{\rm obs}$ is the number
of observations and `Moon illum.' is the fractional illumination of the Moon at the midpoint of the transit.}
\begin{tabular}{llccccccccc} \hline
Transit & Date & Start time & End time &$N_{\rm obs}$& Exposure & Filter & Airmass & Moon & Aperture   & Scatter \\
        &      &    (UT)    &   (UT)   &             & time (s) &        &         &illum.& sizes (px) & (mmag)  \\
\hline
1 & 2009 06 03 & 05:44 & 10:35 & 123 & 90--100 & $R_C$ & 1.66 $\to$ 1.23 $\to$ 1.48 & 0.816 & 26, 33, 100& 0.422 \\
2 & 2009 08 11 & 02:27 & 05:06 &  97 & 50.0    & $R_C$ & 1.34 $\to$ 1.23 $\to$ 1.30 & 0.761 & 23, 40, 70 & --    \\
3 & 2009 08 24 & 01:08 & 05:07 &  83 & 120.0   & $R_C$ & 1.42 $\to$ 1.23 $\to$ 1.45 & 0.179 & 30, 35, 85 & 0.473 \\
4 & 2009 11 11 & 17:00 & 20:11 &  63 & 130.0   & $R_C$ & 1.28 $\to$ 2.17            & 0.273 & 20, 30, 45 & 0.730 \\
\hline \end{tabular} \end{table*}

Aside from our own Solar system, currently the only way to measure the mass and radius of a planet is to study one which transits its parent star. Almost all known transiting extrasolar planets (TEPs) are gas giants with radii similar to Jupiter's but who  orbit much closer to their parent stars. The masses and radii, and thus surface gravities and mean densities, of these objects can be determined to high precision, yielding in turn constraints on their chemical composition and internal structure \citep{Bodenheimer++03apj,Baraffe++08aa,Fortney++07apj,Fortney+08apj}. The flux ratio of a planet to its parent star can also be obtained through measurements of the depth of its occultation (secondary eclipse) at a range of wavelengths, delivering constraints on the structure of extended and highly irradiated atmospheres \citep[e.g.][]{Knutson+08apj,Anderson+10xxx}.

To measure the physical properties of a TEP we need a good light curve covering one or more transits, radial velocity (or astrometric) observations of its parent star at a range of orbital phases, and one additional constraint usually supplied by forcing the properties of the parent star to match predictions from theoretical stellar models. Detailed error budgets of the analysis process \citep{Me09mn} show that the dominant contributor to the uncertainties is almost always the light curve. We have therefore embarked on a project to obtain high-quality light curves of transiting systems and thus measure their physical properties as accurately and homogeneously as possible.

Our approach to obtaining transit light curves is to heavily defocus a telescope, spreading the light from each star over thousands of pixels on the CCD, and take images with comparatively long exposure times of up to 120\,s. Because the light from each star is distributed over thousands of pixels, it is possible to achieve both a very low Poisson noise and minimal flat-fielding errors. Using long exposure times mean we spend less time overall reading out the CCD, so are able to detect more light from the star and decrease the noise due to scintillation. Defocussing the telescope means our observations are subject to a higher background light level, but this is not important for the bright stars around which most known TEPs have been found. In this work we present defocussed observations of the planetary system WASP-2, obtained using two telescopes and achieving scatters ranging from 0.42 to 0.73 mmag per datapoint.

The transiting nature of WASP-2 was discovered by \citet{Cameron+07mn} (see also \citealt{Cameron+07mn2}) in photometry from the SuperWASP-North telescope \citep{Pollacco+06pasp}. The planet WASP-2\,b is slightly larger and less massive than Jupiter, whilst the star WASP-2\,A is rather smaller and cooler than the Sun. Follow-up photometry of one transit was presented and analysed by \citet{Charbonneau+07apj}, and reanalysed by \citet{Torres++08apj} and \citet{Me08mn,Me09mn}. Another transit has also been observed with the William Herschel Telescope \citep{Hrudkova+09iaus}. \citet{Triaud+10aa} has presented a study of WASP-2 which incorporates spectroscopic observations of the Rossiter-McLaughlin effect, spectroscopic estimates of the effective temperature and metal abundances of the host star, and a re-analysis of the light curves from \citet{Charbonneau+07apj} and \citet{Hrudkova+09iaus}. \citet{Wheatley+10} has recently presented infrared {\it Spitzer} observations of the secondary eclipse of the system.

The study of WASP-2 presents an additional complication in the recently-discovered presence of a faint companion star to the WASP-2 system. \citet{Daemgen+09aa} used the high-speed ``lucky-imaging'' camera AstraLux to obtain images of the field of WASP-2 in the SDSS $i$ and $z$ passbands and with a high spatial resolution. These authors discovered a faint point source within 0.76\as\ of WASP-2, and measured magnitude differences of $\Delta i = 4.095 \pm 0.025$ and $\Delta z = 3.626 \pm 0.016$ mag with respect to our target star. These correspond to a light ratio of $0.0354 \pm 0.0005$ and $0.0230 \pm 0.0005$ in the two filters, respectively. Because this faint star was within the point spread function (PSF) of all previous photometric observations, studies which relied on these data will be systematically incorrect by a small amount. In Section\,\ref{sec:lc:l3} we outline our method for accounting for the light from this faint star.


\section{Observations and data reduction}

We observed three transits of WASP-2 in 2009 June and August, using the 1.54\,m Danish Telescope%
\footnote{For information on the 1.54\,m Danish Telescope and DFOSC see:
{\scriptsize\tt http://www.eso.org/sci/facilities/lasilla/telescopes/d1p5/}}
at ESO La Silla, Chile. We also observed one transit on the night of 2009 November 11 using the 1.52\,m G.\ D.\ Cassini Telescope%
\footnote{Information on the 1.52\,m Cassini Telescope and BFOSC can be found at
{\tt http://www.bo.astro.it/loiano/}}
at Loiano Observatory, Italy. The second of our transit observations was strongly affected by cloud, so we do not include these data in our analysis. The fourth transit (Cassini) was also taken in non-photometric conditions, causing the observations to be noisier than intended. An observing log is given in Table\,\ref{tab:obslog}.

Observations taken with the Danish telescope used the DFOSC focal-reducing imager, which has a plate scale of 0.39\as\,px$^{-1}$. The CCD was not binned, but was windowed down by varying amounts to decrease the readout time. All observations were done through a Cousins $R$ filter, and the amount of defocussing was adjusted until the peak counts per pixel from WASP-2 were roughly 25\,000 above the sky background, resulting in doughnut-shaped PSFs with diameters ranging from 30 to 45 pixels. The pointing of the telescope was maintained using autoguiding, and we did not change the amount of defocussing during individual observing sequences. The observing sequence from the Cassini telescope used the BFOSC focal-reducing imager (plate scale 0.58\as\,px$^{-1}$) and the same approach as for the Danish telescope.

\begin{figure} \includegraphics[width=0.48\textwidth,angle=0]{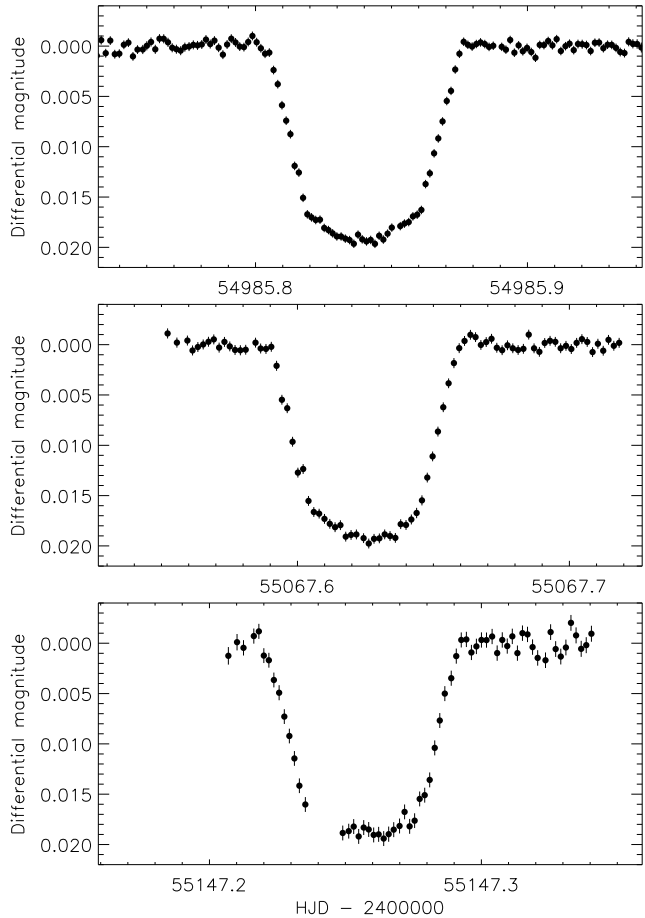}
\caption{\label{fig:plotlc} Light curves of WASP-2 from the three nights of
observations included in our analysis. The error bars have been scaled to
give $\chi^2_{\ \nu} = 1.0$ for each night, and some are smaller than the
symbol sizes.} \end{figure}

\begin{table} \centering \caption{\label{tab:lc} Excerpts of the
light curve of WASP-2. The full dataset will be made available as
a supplement to this paper and at the CDS.}
\begin{tabular}{lccr} \hline
Telescope & HJD & Diff.\ mag. & Uncertainty \\
\hline
Danish  & 2454985.742452 & \ $0.0009568$ & $0.0004018$ \\
Danish  & 2454985.944273 & $-0.0003249$  & $0.0004759$ \\[2pt]
Danish  & 2455054.609274 & \ $0.0025309$ & $0.0005716$ \\
Danish  & 2455054.718127 & \ $0.0246546$ & $0.0016991$ \\[2pt]
Danish  & 2455067.553072 & $-0.0011116$  & $0.0003926$ \\
Danish  & 2455067.719132 & $-0.0001831$  & $0.0003860$ \\[2pt]
Cassini & 2455147.207773 & \ $0.0012495$ & $0.0006328$ \\
Cassini & 2455147.341286 & $-0.0009452$  & $0.0005788$ \\
\hline \end{tabular} \end{table}

Several images were taken with the telescopes properly focussed, to verify that there were no faint stars within the defocussed PSF of WASP-2. The closest detectable star is 4.7\,mag fainter and at a distance of 18\as, and its PSF did not overlap with our target star. We conclude that no stars interfere with the PSF of WASP-2 aside from the one at 0.76\as\ discovered by \citet{Daemgen+09aa}.

Data reduction was performed as in previous papers \citep{Me+09mn,Me+09mn2,Me+09apj}, using a custom pipeline written in the {\sc idl}\footnote{The acronym {\sc idl} stands for Interactive Data Language and is a trademark of ITT Visual
Information Solutions. For further details see {\tt http://www.ittvis.com/ProductServices/IDL.aspx}.} programming language and accessing the {\sc daophot} package \citep{Stetson87pasp} to perform aperture photometry with the {\sc aper} \footnote{{\sc aper} is part of the {\sc astrolib} subroutine library distributed by NASA. For further details see {\tt http://idlastro.gsfc.nasa.gov/}.} routine. The apertures were placed by eye. For transits 1 and 3 their positions were held fixed throughout the observing sequence. For transits 2 and 4 the pointing was monitored by cross-correlating each image with a reference image, and the apertures shifted to follow the stars. We tried a wide range of aperture sizes and retained those which gave photometry with the lowest scatter; in each case we find that the shape of the light curve is very insensitive to the aperture sizes.

Up to 11 comparison stars were measured on each image, and checked for short-period variability. We then selected the five best and combined them into a weighted ensemble which minimised the scatter in observations before and after transit. Simultaneously with this procedure we fitted a straight line to the data outside transit to normalise the light curve and remove slow magnitude drifts due to atmospheric effects.

We also tried flat-fielding and debiassing the observations before measuring aperture photometry. We found that our treatment of the CCD bias level had a negligible effect, and that flat-fielding the data actually made things worse. We therefore debiassed the data but did not flat-field them. The final light curves are shown in Fig.\,\ref{fig:plotlc} and tabulated in Table\,\ref{tab:lc}. The scatter in the final light curves varies from 0.42 to 0.73 mmag. To our knowledge the first transit dataset represents the most precise light curve of a point source ever obtained from a ground-based telescope.


\section{Light curve analysis}                                                                           \label{sec:lc}

The analysis of our light curves was performed using the approach discussed in detail by \citet{Me08mn,Me09mn,Me10mn}, so our results are consistent with the homogeneous analysis of TEPs presented in those works. In short, the light curves were modelled using the {\sc jktebop}\footnote{{\sc jktebop} is written in {\sc fortran77} and the source code is available at {\tt http://www.astro.keele.ac.uk/$\sim$jkt/}} code \citep{Me++04mn,Me++04mn2}, which is based on the {\sc ebop} program \citep{PopperEtzel81aj,Etzel81conf,NelsonDavis72apj} where the two components of a binary system are simulated using biaxial spheroids. The asphericity of the components is governed by the mass ratio, for which we used the value 0.0009. Large changes in this value have a negligible effect on our results, partly because in a system with a circular orbit the times of eclipse are when the sky projections of the components suffer the smallest distortion.

\subsection{Period determination}                                                                   \label{sec:lc:porb}

\begin{table} \begin{center}
\caption{\label{tab:minima} Times of minimum light of WASP-2
and their residuals versus the ephemeris derived in this work.
\newline {\bf References:}     (1) \citet{Cameron+07mn};
(2) \citet{Charbonneau+07apj}; (3) \citet{Hrudkova+09iaus}.}
\begin{tabular}{l@{\,$\pm$\,}l r r l} \hline
\multicolumn{2}{l}{Time of minimum}   & Cycle  & Residual & Reference \\
\multicolumn{2}{l}{(HJD $-$ 2400000)} & no.    & (HJD)    &           \\
\hline
53991.5146  & 0.0044  &     0.0 & $ $0.00005 & 1         \\
54008.73205 & 0.00028 &     8.0 & $-$0.00027 & 2         \\
54357.39254 & 0.00016 &   170.0 & $ $0.00034 & 3         \\
54985.84071 & 0.00009 &   462.0 & $-$0.00015 & This work \\
55067.62554 & 0.00011 &   500.0 & $ $0.00027 & This work \\
55147.25721 & 0.00017 &   537.0 & $-$0.00025 & This work \\[2pt]
54260.5420  & 0.0010  &   125.0 & $-$0.00023 & Gregorio   (AXA)    \\
54277.7580  & 0.0020  &   133.0 & $-$0.00200 & Gary       (AXA)    \\
54279.9110  & 0.0010  &   134.0 & $-$0.00122 & Foote      (AXA)    \\
54316.4991  & 0.0015  &   151.0 & $-$0.00089 & Vanmunster (AXA)    \\
54320.8053  & 0.0010  &   153.0 & $ $0.00087 & Foote      (AXA)    \\
54333.7186  & 0.0020  &   159.0 & $ $0.00084 & Sheridan   (AXA)    \\
54348.7760  & 0.0020  &   166.0 & $-$0.00731 & Sheridan   (AXA)    \\
54357.3906  & 0.0010  &   170.0 & $-$0.00160 & Poddan\'y  (AXA)    \\
54357.3914  & 0.0010  &   170.0 & $-$0.00080 & Ohlert     (AXA)    \\
54710.35882 & 0.00100 &   334.0 & $ $0.00231 & Koci\'an   (TRESCA) \\
54757.7052  & 0.0006  &   356.0 & $-$0.00018 & Gary       (AXA)    \\
54953.55627 & 0.00083 &   447.0 & $-$0.00126 & Br\'at     (TRESCA) \\
54981.53709 & 0.00044 &   460.0 & $ $0.00068 & Sauer      (TRESCA) \\
55009.5168  & 0.0018  &   473.0 & $ $0.00151 & Gonzalez   (AXA)    \\
55024.5839  & 0.0020  &   480.0 & $ $0.00306 & Lobao      (AXA)    \\
55026.7327  & 0.0006  &   481.0 & $-$0.00036 & Norby      (AXA)    \\
55037.4900  & 0.0014  &   486.0 & $-$0.00417 & Salom      (AXA)    \\
55037.4945  & 0.0010  &   486.0 & $ $0.00033 & Naves      (AXA)    \\
55065.4735  & 0.0008  &   499.0 & $ $0.00045 & Gregorio   (AXA)    \\
55069.7799  & 0.0015  &   501.0 & $ $0.00241 & Garlitz    (AXA)    \\
55074.08138 & 0.00050 &   503.0 & $-$0.00055 & Cao        (TRESCA) \\
55095.6050  & 0.0009  &   513.0 & $ $0.00085 & Cordiale   (AXA)    \\
\hline \end{tabular} \end{center} \end{table}

\begin{figure*} \includegraphics[width=\textwidth,angle=0]{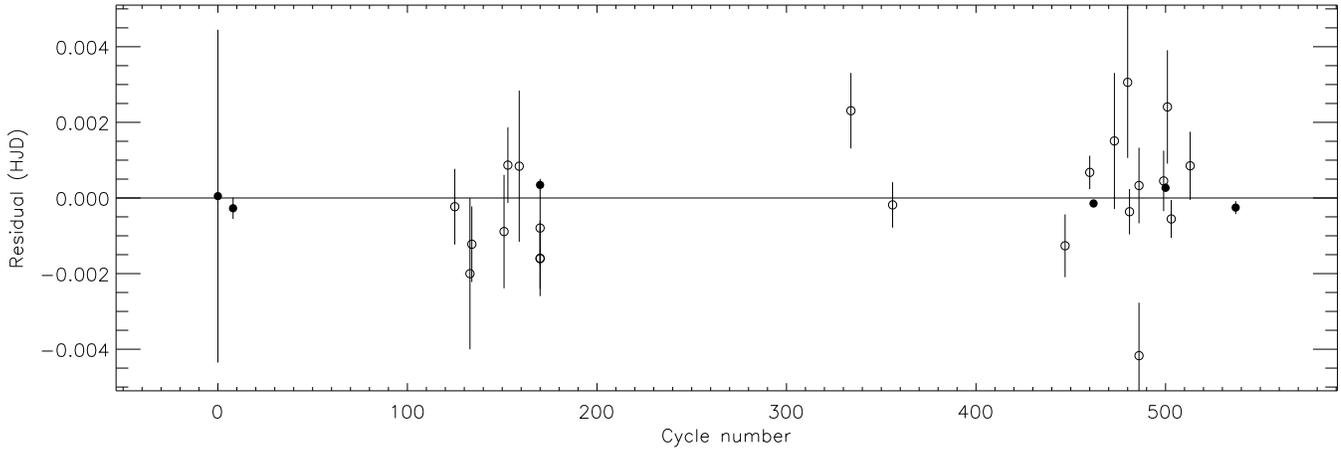}
\caption{\label{fig:minima} Plot of the residuals of the timings of
mid-transit of WASP-2 versus a linear ephemeris. Some error bars
are smaller than the symbol sizes. The timings obtained by amateur
astronomers are plotted using open circles.} \end{figure*}

As a first step in the analysis of WASP-2 we have performed a fit to each of our three light curves separately using {\sc jktebop}. The measurement errors of our photometry (which come from the {\sc aper} routine) were found to be underestimated so were rescaled to give a reduced $\chi^2$ of $\chi^2_\nu = 1.0$ with respect to the best-fitting model \citep[see][]{Bruntt+06aa}.

These individual fits also yielded one time of minimum light each. The uncertainties in these values were estimated using Monte Carlo simulations. We have augmented them with three times of minimum light taken from \citet{Cameron+07mn}, \citet{Charbonneau+07apj} and \citet{Hrudkova+09iaus}, and with an additional 22 minimum timings obtained by amateur astronomers and made available on the AXA\footnote{Amateur Exoplanet Archive, {\tt http://brucegary.net/AXA/x.htm}} and TRESCA\footnote{The TRansiting ExoplanetS and CAndidates (TRESCA) website can be found at, {\tt http://var2.astro.cz/EN/tresca/index.php}} websites. A small number of AXA and TRESCA timings were rejected due to noisy data or incomplete coverage of the transit.

Armed with these times of minimum light we have redetermined the orbital ephemeris of the WASP-2 system to be
$$ T_0 = {\rm HJD} \,\, 2\,453\,991.51455 (17) \, + \, 2.15222144 (39) \times E $$
where $E$ is the number of orbital cycles after the reference epoch and quantities in parentheses denote the uncertainty in the final digit of the preceding number. The $\chi^2_\nu$ of the straight line fitted to the timings is equal to 1.53, so the ephemeris uncertainties were multiplied by the square-root of this number to yield the values quoted above. A plot of the residuals around the fit is shown in Fig.\,\ref{fig:minima}. Including the 22 amateur epochs roughly doubles the  precision of the ephemeris.

There were several cases during the 2009 MiNDSTEp observing season, which is primarily oriented to studying microlensing events towards the Galactic bulge, where photometric observations appeared to occur slightly earlier than expected. The timestamps imprinted in the DFOSC fits file headers are currently under investigation. The Julian Dates for the DFOSC observations presented in the current work should therefore be treated with caution, as should the observations of WASP-18 studied by \citet{Me+09apj}. One of the reasons for obtaining a further transit observation of WASP-2 with the Cassini telescope was to improve its orbital ephemeris and allow an external check on the DFOSC timings. We find that the two DFOSC transits occur $12 \pm 8$\,s {\em later} than would be expected, implying that there is no problem with their timestamps. We have therefore included these timings in our period study.

\subsection{Accounting for the light from the faint star}                                             \label{sec:lc:l3}

Light from the faint star, discovered by \citet{Daemgen+09aa} 0.76\as\ away from WASP-2, must be accounted for to avoid systematically underestimating the radius of the planet. The PSF of WASP-2 in our observations includes essentially all the light from the faint star, and even our focussed observations were unable to pick this star up due to its faintness and closeness to our target star.

\citet{Daemgen+09aa} obtained light ratios between the WASP-2 system and the faint star in the SDSS $i$ and $z$ bands. We have extrapolated these light ratios to the $R_C$ band using synthetic spectra calculated with the {\sc atlas9} model atmospheres \citep{Kurucz79apjs,Kurucz93} and filter response functions taken from the Isaac Newton Group website\footnote{Isaac Newton Group: {\tt www.ing.iac.es}}. We find an effective temperature for the faint star of $\Teff = 3780 \pm 46$\,K and a light ratio of $\ell_{\rm R} = 0.0176 \pm 0.0007$, where the errorbars include contributions from uncertainties in the $i$ and $z$ light ratios and in the \Teff\ of WASP-2\,A. There will be a systematic error in $\ell_{\rm R}$ from the use of theoretical model atmosphere calculations, but this will be very small as the effective wavelength of $R_C$ is close to that of $i$. The systematic error in the \Teff\ of the faint star will be substantially larger, but can be ignored because we do not use this \Teff\ measurement in our analysis.


The light from the faint star can be straightforwardly included in our photometric analysis, as {\sc jktebop} (and other light curve codes) allow for this possibility under the moniker `third light' (symbolised by $L_3$). \citet{Me10mn} modified {\sc jktebop} to allow $L_3$ to be included as a fitted parameter constrained by a measured value. For our defocussed observations, which were all taken through the $R_C$ filter, we adopt $L_3 = 0.0176 \pm 0.0007$. The observations of \citet{Charbonneau+07apj} were obtained using an SDSS $z$ filter, for which a magnitude difference (light ratio) was given directly by \citet{Daemgen+09aa}.

\subsection{Light curve modelling}                                                                    \label{sec:lc:lc}

\begin{table*} \caption{\label{tab:lc:dk} Parameters of the {\sc jktebop} best fits of the
Danish telescope light curve, using different approaches to LD. For each part of the table
the upper quantities are fitted parameters and the lower quantities are derived parameters.
$T_0$ is given as HJD $-$ 2454000.0. The light curve contains 211 datapoints.}
\begin{tabular}{l r@{\,$\pm$\,}l r@{\,$\pm$\,}l r@{\,$\pm$\,}l r@{\,$\pm$\,}l r@{\,$\pm$\,}l}
\hline 
\                     &      \mc{Linear LD law}     &    \mc{Quadratic LD law}    &   \mc{Square-root LD law}   &   \mc{Logarithmic LD law}   &     \mc{Cubic LD law}       \\
\hline
\multicolumn{11}{l}{All LD coefficients fixed:} \\
$r_{\rm A}+r_{\rm b}$ & 0.1391       & 0.0016       & 0.1387       & 0.0016       & 0.1392       & 0.0016       & 0.1394       & 0.0016       & 0.1424       & 0.0014       \\
$k$                   & 0.13342      & 0.00054      & 0.13217      & 0.00058      & 0.13269      & 0.00056      & 0.13249      & 0.00056      & 0.13338      & 0.00041      \\
$i$ (deg.)            & 84.880       &  0.112       & 84.960       &  0.115       & 84.902       &  0.107       & 84.886       &  0.106       & 84.568       &  0.096       \\
$u_{\rm A}$           &      \mc{ 0.60 fixed}       &      \mc{ 0.50 fixed}       &      \mc{ 0.35 fixed}       &      \mc{ 0.72 fixed}       &      \mc{ 0.40 fixed}       \\
$v_{\rm A}$           &            \mc{ }           &      \mc{ 0.20 fixed}       &      \mc{ 0.42 fixed}       &      \mc{ 0.21 fixed}       &      \mc{ 0.10 fixed}       \\
$T_0$                 & 985.840869   &   0.000071   & 985.840868   &   0.000072   & 985.840868   &   0.000073   & 985.840868   &   0.000071   & 985.840869   &   0.000067   \\[2pt]
$r_{\rm A}$           & 0.1227       & 0.0014       & 0.1225       & 0.0014       & 0.1229       & 0.0013       & 0.1231       & 0.0014       & 0.1256       & 0.0012       \\
$r_{\rm b}$           & 0.01637      & 0.00024      & 0.01619      & 0.00025      & 0.01631      & 0.00023      & 0.01631      & 0.00024      & 0.01676      & 0.00020      \\
$\sigma$ ($m$mag)     &        \mc{ 0.4833}         &        \mc{ 0.4857}         &        \mc{ 0.4843}         &        \mc{ 0.4843}         &        \mc{ 0.4919}         \\
$\chi^2_\nu$          &        \mc{ 1.1452}         &        \mc{ 1.1580}         &        \mc{ 1.1505}         &        \mc{ 1.1505}         &        \mc{ 1.1874}         \\
\hline
\multicolumn{11}{l}{Fitting for the linear LD coefficient and perturbing the nonlinear LD coefficient:} \\
$r_{\rm A}+r_{\rm b}$ & 0.1393       & 0.0019       & 0.1397       & 0.0020       & 0.1398       & 0.0020       & 0.1398       & 0.0020       & 0.1398       & 0.0020       \\
$k$                   & 0.13350      & 0.00066      & 0.13259      & 0.00068      & 0.13293      & 0.00064      & 0.13262      & 0.00071      & 0.13299      & 0.00071      \\
$i$ (deg.)            & 84.85        &  0.15        & 84.84        &  0.16        & 84.83        &  0.16        & 84.84        &  0.16        & 84.83        &  0.16        \\
$u_{\rm A}$           & 0.585        & 0.043        & 0.437        & 0.070        & 0.312        & 0.061        & 0.692        & 0.062        & 0.534        & 0.055        \\
$v_{\rm A}$           &            \mc{ }           &     \mc{ 0.20 perturbed}    &     \mc{ 0.42 perturbed}    &     \mc{ 0.21 perturbed}    &     \mc{ 0.10 perturbed}    \\
$T_0$                 & 985.840869   &   0.000067   & 985.840868   &   0.000071   & 985.840869   &   0.000070   & 985.840869   &   0.000073   & 985.840869   &   0.000071   \\[2pt]
$r_{\rm A}$           & 0.1229       & 0.0017       & 0.1234       & 0.0017       & 0.1234       & 0.0017       & 0.1234       & 0.0017       & 0.1234       & 0.0018       \\
$r_{\rm b}$           & 0.01641      & 0.00029      & 0.01636      & 0.00029      & 0.01641      & 0.00030      & 0.01637      & 0.00030      & 0.01640      & 0.00030      \\
$\sigma$ ($m$mag)     &        \mc{ 0.4831}         &        \mc{ 0.4839}         &        \mc{ 0.4837}         &        \mc{ 0.4840}         &        \mc{ 0.4837}         \\
$\chi^2_\nu$          &        \mc{ 1.1499}         &        \mc{ 1.1539}         &        \mc{ 1.1529}         &        \mc{ 1.1541}         &        \mc{ 1.1528}         \\
\hline 
\end{tabular} \end{table*}

\begin{table*} \caption{\label{tab:lc:lo} As Table\,\ref{tab:lc:dk} but for the Cassini data.
$T_0$ is given as HJD $-$ 2455000.0 and the light curve contains 63 datapoints.}
\begin{tabular}{l r@{\,$\pm$\,}l r@{\,$\pm$\,}l r@{\,$\pm$\,}l r@{\,$\pm$\,}l r@{\,$\pm$\,}l}
\hline 
\                     &      \mc{Linear LD law}     &    \mc{Quadratic LD law}    &   \mc{Square-root LD law}   &   \mc{Logarithmic LD law}   &     \mc{Cubic LD law}       \\
\hline
\multicolumn{11}{l}{All LD coefficients fixed:} \\
$r_{\rm A}+r_{\rm b}$ & 0.1382       & 0.0041       & 0.1371       & 0.0046       & 0.1379       & 0.0042       & 0.1380       & 0.0044       & 0.1415       & 0.0035       \\
$k$                   & 0.1320       & 0.0015       & 0.1305       & 0.0017       & 0.1311       & 0.0016       & 0.1309       & 0.0016       & 0.1324       & 0.0011       \\
$i$ (deg.)            & 85.01        &  0.29        & 85.15        &  0.32        & 85.07        &  0.30        & 85.05        &  0.31        & 84.69        &  0.23        \\
$u_{\rm A}$           &      \mc{ 0.60 fixed}       &      \mc{ 0.50 fixed}       &      \mc{ 0.35 fixed}       &      \mc{ 0.72 fixed}       &      \mc{ 0.40 fixed}       \\
$v_{\rm A}$           &            \mc{ }           &      \mc{ 0.20 fixed}       &      \mc{ 0.42 fixed}       &      \mc{ 0.21 fixed}       &      \mc{ 0.10 fixed}       \\
$T_0$                 & 147.25724    &   0.00017    & 147.25725    &   0.00018    & 147.25724    &   0.00017    & 147.25724    &   0.00018    & 147.25721    &   0.00017    \\[2pt]
$r_{\rm A}$           & 0.1221       & 0.0035       & 0.1212       & 0.0038       & 0.1219       & 0.0036       & 0.1221       & 0.0038       & 0.1249       & 0.0030       \\
$r_{\rm b}$           & 0.01611      & 0.00062      & 0.01582      & 0.00070      & 0.01598      & 0.00063      & 0.01598      & 0.00066      & 0.01654      & 0.00050      \\
$\sigma$ ($m$mag)     &        \mc{ 0.7385}         &        \mc{ 0.7417}         &        \mc{ 0.7387}         &        \mc{ 0.7371}         &        \mc{ 0.7260}         \\
$\chi^2_\nu$          &        \mc{ 1.0248}         &        \mc{ 1.0344}         &        \mc{ 1.0255}         &        \mc{ 1.0206}         &        \mc{ 0.9875}         \\
\hline
\multicolumn{11}{l}{Fitting for the linear LD coefficient and perturbing the nonlinear LD coefficient:} \\
$r_{\rm A}+r_{\rm b}$ & 0.1426       & 0.0049       & 0.1427       & 0.0052       & 0.1427       & 0.0051       & 0.1427       & 0.0049       & 0.1427       & 0.0050       \\
$k$                   & 0.1330       & 0.0014       & 0.1321       & 0.0014       & 0.1324       & 0.0014       & 0.1322       & 0.0013       & 0.1325       & 0.0014       \\
$i$ (deg.)            & 84.58        &  0.41        & 84.58        &  0.42        & 84.58        &  0.41        & 84.58        &  0.38        & 84.58        &  0.41        \\
$u_{\rm A}$           & 0.39         & 0.18         & 0.23         & 0.21         & 0.12         & 0.19         & 0.49         & 0.19         & 0.34         & 0.20         \\
$v_{\rm A}$           &            \mc{ }           &     \mc{ 0.20 perturbed}    &     \mc{ 0.42 perturbed}    &     \mc{ 0.21 perturbed}    &     \mc{ 0.10 perturbed}    \\
$T_0$                 & 147.25721    &   0.00018    & 147.25721    &   0.00017    & 147.25721    &   0.00017    & 147.25721    &   0.00017    & 147.25721    &   0.00017    \\[2pt]
$r_{\rm A}$           & 0.1258       & 0.0042       & 0.1261       & 0.0045       & 0.1260       & 0.0044       & 0.1260       & 0.0042       & 0.1260       & 0.0043       \\
$r_{\rm b}$           & 0.01673      & 0.00063      & 0.01666      & 0.00068      & 0.01668      & 0.00065      & 0.01666      & 0.00065      & 0.01669      & 0.00064      \\
$\sigma$ ($m$mag)     &        \mc{ 0.7251}         &        \mc{ 0.7245}         &        \mc{ 0.7246}         &        \mc{ 0.7245}         &        \mc{ 0.7246}         \\
$\chi^2_\nu$          &        \mc{ 1.0021}         &        \mc{ 1.0003}         &        \mc{ 1.0007}         &        \mc{ 1.0002}         &        \mc{ 1.0007}         \\
\hline 
\end{tabular} \end{table*}

\begin{table*} \caption{\label{tab:lc:ch} As Table\,\ref{tab:lc:dk} but for the \citet{Charbonneau+07apj} data.
 $T_0$ is given as HJD $-$ 2454000.0. The light curve contains 426 datapoints.}
\begin{tabular}{l r@{\,$\pm$\,}l r@{\,$\pm$\,}l r@{\,$\pm$\,}l r@{\,$\pm$\,}l r@{\,$\pm$\,}l}
\hline 
\                     &      \mc{Linear LD law}     &    \mc{Quadratic LD law}    &   \mc{Square-root LD law}   &   \mc{Logarithmic LD law}   &     \mc{Cubic LD law}       \\
\hline
\multicolumn{11}{l}{All LD coefficients fixed:} \\
$r_{\rm A}+r_{\rm b}$ & 0.1390       & 0.0049       & 0.1397       & 0.0045       & 0.1399       & 0.0044       & 0.1399       & 0.0047       & 0.1416       & 0.0041       \\
$k$                   & 0.1343       & 0.0016       & 0.1331       & 0.0014       & 0.1337       & 0.0013       & 0.1334       & 0.0014       & 0.1340       & 0.0011       \\
$i$ (deg.)            & 85.01        &  0.35        & 84.97        &  0.31        & 84.92        &  0.31        & 84.93        &  0.33        & 84.73        &  0.28        \\
$u_{\rm A}$           &      \mc{ 0.54 fixed}       &      \mc{ 0.29 fixed}       &      \mc{ 0.13 fixed}       &      \mc{ 0.61 fixed}       &      \mc{ 0.30 fixed}       \\
$v_{\rm A}$           &            \mc{ }           &      \mc{ 0.30 fixed}       &      \mc{ 0.53 fixed}       &      \mc{ 0.26 fixed}       &      \mc{ 0.10 fixed}       \\
$T_0$                 & 8.73204      & 0.00021      & 8.73204      & 0.00022      & 8.73205      & 0.00020      & 8.73205      & 0.00022      & 8.73205      & 0.00020      \\[2pt]
$r_{\rm A}$           & 0.1225       & 0.0042       & 0.1233       & 0.0039       & 0.1234       & 0.0038       & 0.1234       & 0.0040       & 0.1249       & 0.0036       \\
$r_{\rm b}$           & 0.01646      & 0.00072      & 0.01641      & 0.00064      & 0.01649      & 0.00064      & 0.01646      & 0.00067      & 0.01674      & 0.00057      \\
$\sigma$ ($m$mag)     &        \mc{ 1.8878}         &        \mc{ 1.8873}         &        \mc{ 1.8869}         &        \mc{ 1.8869}         &        \mc{ 1.8891}         \\
$\chi^2_\nu$          &        \mc{ 0.9975}         &        \mc{ 0.9970}         &        \mc{ 0.9965}         &        \mc{ 0.9966}         &        \mc{ 0.9989}         \\
\hline
\multicolumn{11}{l}{Fitting for the linear LD coefficient and perturbing the nonlinear LD coefficient:} \\
$r_{\rm A}+r_{\rm b}$ & 0.1395       & 0.0057       & 0.1400       & 0.0060       & 0.1399       & 0.0055       & 0.1400       & 0.0060       & 0.1397       & 0.0058       \\
$k$                   & 0.1344       & 0.0016       & 0.1332       & 0.0014       & 0.1337       & 0.0015       & 0.1335       & 0.0015       & 0.1339       & 0.0016       \\
$i$ (deg.)            & 84.93        &  0.46        & 84.91        &  0.46        & 84.92        &  0.45        & 84.91        &  0.47        & 84.93        &  0.47        \\
$u_{\rm A}$           & 0.47         & 0.15         & 0.24         & 0.17         & 0.13         & 0.16         & 0.60         & 0.16         & 0.42         & 0.16         \\
$v_{\rm A}$           &            \mc{ }           &     \mc{ 0.30 perturbed}    &     \mc{ 0.53 perturbed}    &     \mc{ 0.26 perturbed}    &     \mc{ 0.10 perturbed}    \\
$T_0$                 & 8.73204      & 0.00022      & 8.73205      & 0.00022      & 8.73205      & 0.00021      & 8.73205      & 0.00022      & 8.73205      & 0.00021      \\[2pt]
$r_{\rm A}$           & 0.1229       & 0.0049       & 0.1236       & 0.0051       & 0.1234       & 0.0048       & 0.1236       & 0.0052       & 0.1232       & 0.0050       \\
$r_{\rm b}$           & 0.01652      & 0.00081      & 0.01646      & 0.00078      & 0.01650      & 0.00077      & 0.01649      & 0.00083      & 0.01650      & 0.00081      \\
$\sigma$ ($m$mag)     &        \mc{ 1.8869}         &        \mc{ 1.8869}         &        \mc{ 1.8869}         &        \mc{ 1.8869}         &        \mc{ 1.8869}         \\
$\chi^2_\nu$          &        \mc{ 0.9988}         &        \mc{ 0.9989}         &        \mc{ 0.9989}         &        \mc{ 0.9989}         &        \mc{ 0.9988}         \\
\hline 
\end{tabular} \end{table*}

\begin{figure} \includegraphics[width=0.48\textwidth,angle=0]{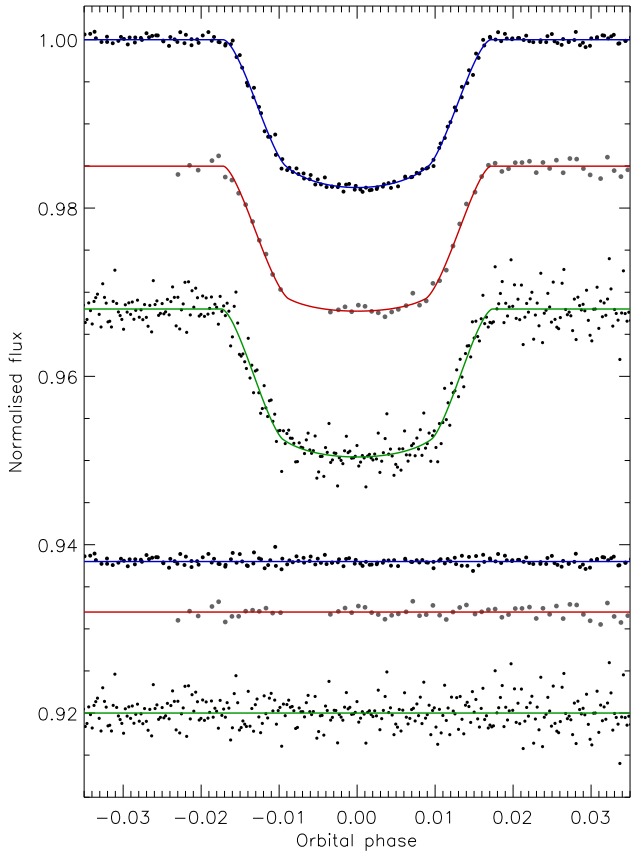}
\caption{\label{fig:lc:fit} Phased light curves of WASP-2 from the Danish telescope
(top), the Cassini telescope (middle grey points) and from \citet{Charbonneau+07apj}
(lower), compared to the best fits found using {\sc jktebop} and the quadratic LD
law. The residuals of the fits are plotted at the bottom of the figure, offset from
zero.} \end{figure}

Our light curves have been modelled using the {\sc jktebop} code. The main parameters of the model are the fractional radii of the star and planet, $r_{\rm A}$ and $r_{\rm b}$ (the radii of the components divided by the semimajor axis), and the orbital inclination, $i$. \ $r_{\rm A}$ and $r_{\rm b}$ are actually incorporated as their sum, $r_{\rm A}$$+$$r_{\rm b}$, and ratio, $k$, as these parameters are more weakly correlated.

The two Danish telescope datasets were solved simultaneously and the Cassini telescope data were studied separately. We have also revisited the $z$-band light curve presented by \citet{Charbonneau+07apj}, and this solution supersedes that of \citet{Me08mn} which was presented before the faint star nearby WASP-2 was discovered.

Limb darkening (LD) was included using any of five LD laws (see \citealt{Me08mn}) and with theoretical coefficients ($u_{\rm A}$ and $v_{\rm A}$) obtained by bilinear interpolation in the tables of \citet{Vanhamme93aj}, \citet{Claret00aa,Claret04aa2} and \citet{ClaretHauschildt03aa}. Solutions were obtained with $u_{\rm A}$ and $v_{\rm A}$ both fixed, both included as fitted parameters, and with $u_{\rm A}$ fitted and $v_{\rm A}$ perturbed by $\pm$0.10 on a flat distribution. For our final solutions we adopt the last of these three possibilities, as the data cannot support the extraction of both LD coefficients. This does not cause a significant dependence on stellar theory because the two LD coefficients are very strongly correlated \citep{Me++07aa}.

We assessed the uncertainties of the fitted parameters using 1000 Monte Carlo simulations for each solution \citep{Me+04mn3,Me+05mn}. The residual-permutation method \citep{Jenkins++02apj} was also used to account for correlated noise, which was found to be unimportant for all sets of observations. The results for each light curve are given in Tables \ref{tab:lc:dk}, \ref{tab:lc:lo} and \ref{tab:lc:ch}. 

The final value quoted for each light curve parameter is the unweighted mean of the values from the four nonlinear LD law solutions. Its uncertainty is the larger of two possibilities: the largest of the individual Monte Carlo uncertainties with the largest deviation from the mean value added in quadrature; or the uncertainty found by the residual-permutation method. The best fits are shown in Fig.\,\ref{fig:lc:fit} and their parameters are tabulated in Table\,\ref{tab:lc:fit}. The agreement is unusually good, with $\chi^2_\nu$ values of between 0.1 and 0.3 in the final weighted-mean values. Table\,\ref{tab:lc:fitcompare} contains literature values of the light curve parameters; a direct comparison with our results in Table\,\ref{tab:lc:fit} shows not only a good agreement but also a substantial decrease in the errorbars. A quick check of the Monte Carlo simulation output shows that $L_3$ is almost uncorrelated with the other fitting parameters. The inclusion of third light in the light curve solutions does not cause any difficulties in this case because its value is small and well-determined.

\begin{table*} \caption{\label{tab:lc:fit} Parameters of the
fits to the three light curves of WASP-2. The final parameters
are the weighted mean of the result for the three light curves.}
\begin{tabular}{l r@{\,$\pm$\,}l r@{\,$\pm$\,}l r@{\,$\pm$\,}l r@{\,$\pm$\,}l r@{\,$\pm$\,}l}
\hline
\ & \mc{This work (Danish)} & \mc{This work (Cassini)} & \mc{This work (Charbonneau)} & \mc{\bf This work (final)} \\
\hline
$r_{\rm A}+r_{\rm b}$ & 0.1398 & 0.0024 & 0.1427 & 0.0057 & 0.1399 & 0.0067 & 0.1402 & 0.0021 \\
$k$                   & 0.1328 & 0.0008 & 0.1323 & 0.0015 & 0.1336 & 0.0017 & 0.1328 & 0.0007 \\
$i$ ($^\circ$)        & 84.84 & 0.19    & 84.58 & 0.47    & 84.92 & 0.52    & 84.81 & 0.17    \\
$r_{\rm A}$           & 0.1234 & 0.0021 & 0.1260 & 0.0050 & 0.1234 & 0.0058 & 0.1237 & 0.0018 \\
$r_{\rm b}$           & 0.01638&0.00036 & 0.01667&0.00076 & 0.01649&0.00092 & 0.01644&0.00031 \\
\hline \end{tabular} \end{table*}

\begin{table*} \caption{\label{tab:lc:fitcompare} Light curve parameters of
WASP-2 taken from the literature, for comparison to the results presented
in Table\,\ref{tab:lc:fit}. Quantities without uncertainties have been
calculated from other parameters given in that study. Only the study by
\citet{Daemgen+09aa} accounts for the light from the faint companion star.}
\setlength{\tabcolsep}{4pt}
\begin{tabular}{l r@{\,$\pm$\,}l r@{\,$\pm$\,}l r@{\,$\pm$\,}l r@{\,$\pm$\,}l r@{\,$\pm$\,}l r@{\,$\pm$\,}l r@{\,$\pm$\,}l}
\hline
\ & \mc{Collier Cameron} & \mc{Charbonneau} & \mc{Torres et al.} & \mc{Southworth} & \mc{Daemgen+ et al.}
    & \mc{Triaud et al.} & \mc{Wheatley et al.} \\
\ & \mc{et al.\ \citeyearpar{Cameron+07mn}} & \mc{et al.\ \citeyearpar{Charbonneau+07apj}}
    & \mc{\citeyearpar{Torres++08apj}} & \mc{\citeyearpar{Me08mn}} & \mc{\citeyearpar{Daemgen+09aa}}
      & \mc{\citeyearpar{Triaud+10aa}} & \mc{\citeyearpar{Wheatley+10}} \\
\hline
$r_{\rm A}$$+$$r_{\rm b}$& \mc{0.086 to 0.132} & \mc{ }      & \mc{0.1423}                & 0.1409 & 0.0067 & \mc{0.1420}   & \mc{0.14155}                   & \mc{ } \\
$k$                      & \mc{0.119 to 0.140} &0.1309&0.0015&\erc{0.1310}{0.0013}{0.0013}& 0.1313 & 0.0017 & 0.1332&0.0015 & \erc{0.1342}{0.0010}{0.0009}   & 0.1334 & 0.0015 \\
$i$ ($^\circ$)           & \mc{ }              & 84.74&0.39  & \erc{84.81}{0.35}{0.27}    & 84.83 & 0.53    & 84.80 & 0.39  & \erc{84.73}{0.18}{0.19}        & \mc{ } \\
$r_{\rm A}$              & \mc{0.123}          & \mc{ }      &\erc{0.1258}{0.0031}{0.0053}& 0.1245 & 0.0058 & \mc{0.1250}   & \erc{0.1248}{0.0025}{0.0024}   & \mc{ } \\
$r_{\rm b}$              &\mc{0.0105 to 0.0202}& \mc{ }      & \mc{0.01648}               & 0.01635&0.00093 & \mc{0.01701}  & \erc{0.01675}{0.00045}{0.00040}& \mc{ } \\
\hline \end{tabular} \end{table*}

\begin{table*} \caption{\label{tab:model} Estimates of the physical
properties of WASP-2. In each case $g_{\rm b} = 19.30 \pm 0.82$\mss,
$\rho_{\rm A} = 1.531 \pm 0.067$\psun\ and $\Teq = 1280 \pm 21$\,K.}
\begin{tabular}{l r@{\,$\pm$\,}l r@{\,$\pm$\,}l r@{\,$\pm$\,}l r@{\,$\pm$\,}l r@{\,$\pm$\,}l r@{\,$\pm$\,}l}
\hline
\ & \mc{This work} & \mc{This work} & \mc{This work} & \mc{This work} & \mc{This work} & \mc{This work} \\
\ & \mc{(Mass-radius)} & \mc{({\it Claret} models)} & \mc{({\it Y$^2$} models)} & \mc{({\it Teramo} models)} & \mc{({\it VRSS} models)} & \mc{({\it DSEP} models)} \\
\hline
$K_{\rm b}$     (\kms) & 150.9   &   4.3   & 154.7   &   3.0    & 153.2   &   2.6   & 151.0   &   1.9    & 151.2   &   3.0    & 152.8   &   2.2    \\
$M_{\rm A}$    (\Msun) & 0.777   & 0.066   & 0.838   & 0.048    & 0.813   & 0.041   & 0.778   & 0.029    & 0.782   & 0.047    & 0.807   & 0.035    \\
$R_{\rm A}$    (\Rsun) & 0.798   & 0.028   & 0.818   & 0.018    & 0.810   & 0.017   & 0.798   & 0.014    & 0.799   & 0.019    & 0.808   & 0.015    \\
$\log g_{\rm A}$ (cgs) & 4.525   & 0.015   & 4.536   & 0.016    & 4.531   & 0.015   & 4.525   & 0.015    & 4.526   & 0.017    & 4.530   & 0.015    \\[2pt]
$M_{\rm b}$    (\Mjup) & 0.828   & 0.050   & 0.871   & 0.038    & 0.854   & 0.033   & 0.829   & 0.026    & 0.832   & 0.038    & 0.849   & 0.030    \\
$R_{\rm b}$    (\Rjup) & 1.032   & 0.035   & 1.058   & 0.029    & 1.048   & 0.026   & 1.033   & 0.023    & 1.034   & 0.029    & 1.045   & 0.025    \\
$\rho_{\rm b}$ (\pjup) & 0.754   & 0.050   & 0.735   & 0.046    & 0.742   & 0.046   & 0.753   & 0.046    & 0.752   & 0.048    & 0.744   & 0.046    \\
\safronov\             & 0.0619  & 0.0024  & 0.0604  & 0.0020   & 0.0610  & 0.0019  & 0.0619  & 0.0018   & 0.0618  & 0.0021   & 0.0612  & 0.0019   \\[2pt]
$a$               (AU) & 0.03000 & 0.00085 & 0.03076 & 0.00060  & 0.03046 & 0.00051 & 0.03002 & 0.00037  & 0.03006 & 0.00059  & 0.03038 & 0.00044  \\
Age              (Gyr) &       \mc{ }      &\erc{10.1}{8.1}{3.8}&\erc{9.9}{3.8}{3.4}&\erc{15.1}{2.6}{2.3}&\erc{14.8}{2.6}{4.4}&\erc{9.3}{2.6}{2.7} \\
\hline \end{tabular} \end{table*}

\begin{table*} \caption{\label{tab:fmodel} Final physical properties
of the WASP-2 system plus a comparison to literature measurements.}
\setlength{\tabcolsep}{4pt}
\begin{tabular}{l r@{\,$\pm$\,}c@{\,$\pm$\,}l r@{\,$\pm$\,}l r@{\,$\pm$\,}l r@{\,$\pm$\,}c@{\,$\pm$\,}l r@{\,$\pm$\,}l r@{\,$\pm$\,}l}
\hline
\ & \mcc{\bf This work (final)} & \mc{Charbonneau} & \mc{Torres et al.} & \mcc{\citet{Me09mn}} & \mc{Daemgen et al.} & \mc{Triaud et al.} \\
\ & \mcc{ } & \mc{et al. \citeyearpar{Charbonneau+07apj}} & \mc{\citeyearpar{Torres++08apj}}
    & \mcc{ } & \mc{\citeyearpar{Daemgen+09aa}} & \mc{\citeyearpar{Triaud+10aa}}\\
\hline
$M_{\rm A}$    (\Msun) & 0.804   & 0.048   & 0.034      & 0.813 & 0.032 & \erm{0.89}{0.12}{0.12}         & 0.88  & 0.11  & 0.03   & 0.89 & 0.12     & \erc{0.84}{0.11}{0.12}      \\
$R_{\rm A}$    (\Rsun) & 0.807   & 0.019   & 0.011      & 0.81 & 0.04   & \erm{0.840}{0.062}{0.065}      & 0.835 & 0.045 & 0.017  & 0.843 & 0.063   & \erc{0.825}{0.042}{0.040}   \\
$\log g_{\rm A}$ (cgs) & 4.530   & 0.017   & 0.006      & \mc{ }        & \erm{4.537}{0.035}{0.046}      & 4.536 & 0.050 & 0.005  & \mc{ }          & \mc{ }                      \\
$\rho_{\rm A}$ (\psun) & \mcc{$1.531 \pm 0.067$}        & \mc{ }        & \erm{1.45}{0.19}{0.11}         & \mcc{$1.50 \pm 0.21$}  & 1.49 & 0.15     & \erc{1.491}{0.088}{0.085}   \\[2pt]
$M_{\rm b}$    (\Mjup) & 0.847   & 0.038   & 0.024      & \mc{ }        & \erm{0.915}{0.090}{0.093}      & 0.91  & 0.10  & 0.02   & 0.914 & 0.092   & \erc{0.866}{0.076}{0.084}   \\
$R_{\rm b}$    (\Rjup) & 1.044   & 0.029   & 0.015      & 1.038 & 0.050 & \erm{1.071}{0.080}{0.083}      & 1.068 & 0.076 & 0.012  & 1.117 & 0.082   & \erc{1.077}{0.055}{0.058}   \\
$g_{\rm b}$      (\ms) & \mcc{$19.30 \pm 0.82$}         & \mc{ }        & \erm{19.4}{1.8}{1.4}           & \mcc{$19.7 \pm 2.7$}   & 19.0 & 1.6      & \mc{ }                      \\
$\rho_{\rm b}$ (\pjup) & 0.745   & 0.048   & 0.010      & \mc{ }        & \erm{0.74}{0.22}{0.16}         & 0.74  & 0.14  & 0.01   & 0.70 & 0.19     & \mc{ }                      \\[2pt]
\Teq\              (K) & \mcc{$1280 \pm 21$}            & \mc{ }        & \erm{1304}{54}{54}             & \mcc{ }                & 1300 & 54       & \mc{ }                      \\
\safronov\             & 0.0612  & 0.0021  & 0.0009     & \mc{ }        & \erm{0.0592}{0.0046}{0.0041}   & \mcc{ }                & 0.0590 & 0.0044 & \mc{ }                      \\
$a$               (AU) & 0.03034 & 0.00060 & 0.00042    & \mc{ }        & \erm{0.03138}{0.00130}{0.00154}&0.03120&0.00130&0.00030 & 0.03138&0.00142 & \erc{0.0307}{0.0013}{0.0015}\\
Age              (Gyr) &\ermcc{11.8}{8.1}{4.4}{3.3}{2.5}& \mc{ }        & \erm{5.6}{8.4}{5.6}            & \mcc{unconstrained}    & \mc{ }          & \mc{ }                      \\
\hline \end{tabular} \end{table*}

\section{The physical properties of WASP-2}

In order to determine the physical properties we need the light curve parameters determined above, the velocity amplitude of the star ($K_{\rm A} = 153.6 \pm 3.0$\kms; \citealt{Triaud+10aa}), and one additional constraint. We supply this extra constraint by forcing the physical properties of the star to match predictions from theoretical stellar models, guided by the observed effective temperature ($\Teff = 5150 \pm 80$\,K; \citealt{Triaud+10aa}) and metallicity ($\FeH = -0.08 \pm 0.08$; \citealt{Triaud+10aa}). For the parameter which governs the solution process we use the velocity amplitude of the {\em planet}, $K_{\rm b}$.

For the theoretical models we use five different sets of tabulations: {\it Claret} \citep{Claret04aa,Claret05aa,Claret06aa2,Claret07aa2}, {\it Y$^2$} \citep{Demarque+04apjs}, {\it Teramo} \citep{Pietrinferni+04apj}, {\it VRSS} \citep{Vandenberg++06apjs} and {\it DSEP} \citep{Dotter+08apjs}. A detailed description of this solution process can be found in \citet{Me09mn} and a discussion of the five stellar models and the alternative empirical mass--radius relation is given by \citet{Me10mn}. The use of five different model sets allows the assessment of their interagreement and thus the systematic error engendered by relying on theoretical predictions. In reality this will yield only a lower limit on the systematic errors, as the different models have a lot of commonality in their input physics. There are two parameters which do not depend on the stellar model tabulations -- the surface gravity and equilibrium temperature of the planet \citep{Me++07mn,Me10mn} -- and one for which the dependence is negligible -- the stellar density \citep{SeagerMallen03apj}.

The results of the above analysis are estimates of the physical properties of the WASP-2 system, given in Table\,\ref{tab:model} for each stellar model set and for the empirical stellar mass--radius relation defined by \citet{Me08mn}. The mass, radius, surface gravity and density of the star are denoted by $M_{\rm A}$, $R_{\rm A}$, $\log g_{\rm A}$ and $\rho_{\rm A}$, and of the planet by $M_{\rm b}$, $R_{\rm b}$, $g_{\rm b}$ and $\rho_{\rm b}$. The orbital semimajor axis ($a$), equilibrium temperature surrogate
\begin{equation}
\Teq = \Teff \sqrt{\frac{R_{\rm A}}{2a}}
\end{equation}
and \citet{Safronov72} number (\safronov) are also determined.

Table\,\ref{tab:fmodel} contains the final physical properties from this work, calculated as the unweighted mean of the individual results for the five stellar models. The statistical error for each parameter is the largest from the models, and the systematic error the standard deviation of the individual model results. Table\,\ref{tab:fmodel} also contains results from published studies of WASP-2, which are in good agreement with our results but are less precise.


\section{Summary}

Obtaining a high-precision light curve is a crucial part of measuring the physical properties of transiting extrasolar planets. In this work we have presented photometric observations obtained through a heavily defocussed telescope and covering three transits of WASP-2. The first of these observing sequences has a scatter of only 0.42\,mmag around the best-fitting model, which to our knowledge is a new record for a ground-based light curve of planetary transit.

These data were analysed with the {\sc jktebop} code, taking into account the contaminating light from a fainter companion star within 0.76\as\ of the sky position of the WASP-2 system. With the addition of the observed effective temperature, metallicity and velocity amplitude of the host star, plus several sets of theoretical stellar model predictions, we have obtained the physical properties of the system. The planet WASP-2\,b has a mass of $0.847 \pm 0.038 \pm 0.024$\Mjup\ (statistical and systematic errors) and a radius of $1.044 \pm 0.029 \pm 0.015$\Rjup, which is typical for transiting gas giants. Its host star is a 0.8\Msun\ and 0.8\Rsun\ K1 dwarf with a large and largely unconstrained age. The equilibrium temperature of the planet, $1280 \pm 21$\,K, puts it firmly into the pL class \citep{Fortney+08apj}.

The high precision of our mass and radius measurements for WASP-2\,b means this planet is now one of the best-understood TEPs. This is partly due to the small scatter in our light curves, but also to the physical configuration of the system. Transiting planets with relatively low orbital inclinations suffer from less correlation between inclination and the stellar radius, resulting in better-defined light curve solutions. Low-inclination TEPs also spend a larger fraction of their time in the partial phases of their eclipses, making transit timings more accurate. We find uncertainties of as little as 8\,s in the times of the three transits presented here: WASP-2 would be an excellent candidate for a transit timing study, and such data would further improve measurements of the physical properties of this planetary system.


\section*{Acknowledgments}

The reduced light curves presented in this work will be made available at the CDS ({\tt http://cdsweb.u-strasbg.fr/}) and at {\tt http://www.astro.keele.ac.uk/$\sim$jkt/}. JS acknowledges financial support from STFC in the form of a postdoctoral research position under the grant number ST/F002599/1. Astronomical research at the Armagh Observatory is funded by the Northern Ireland Department of Culture, Arts and Leisure (DCAL). DR (boursier FRIA) and J\,Surdej acknowledge support from the Communaut\'e fran\c{c}aise de Belgique - Actions de recherche concert\'ees - Acad\'emie Wallonie-Europe. The following internet-based resources were used in research for this paper: the ESO Digitized Sky Survey; the NASA Astrophysics Data System; the SIMBAD database operated at CDS, Strasbourg, France; and the ar$\chi$iv scientific paper preprint service operated by Cornell University.

\bibliographystyle{mn_new}

\end{document}